\documentclass[]{aastex631}

\newcommand{\HI}{H\,{\sc i}}

\usepackage{graphicx}
\usepackage{amsmath}

\shorttitle{The first FRB host galaxy redshift from radio observations}
\shortauthors{Glowacki et al.}
\graphicspath{{./}{figures/}}

\begin{document}

\title{HI, FRB, what's your z: The first FRB host galaxy redshift from radio observations}

\correspondingauthor{Marcin Glowacki}
\email{marcin.glowacki@curtin.edu.au}

\author[0000-0002-5067-8894]{M. Glowacki}
\affiliation{International Centre for Radio Astronomy Research (ICRAR), Curtin University, Bentley, WA 6102, Australia}

\author{A. Bera}
\affiliation{International Centre for Radio Astronomy Research (ICRAR), Curtin University, Bentley, WA 6102, Australia}

\author[0000-0003-4844-8659]{K. Lee-Waddell}
\affiliation{International Centre for Radio Astronomy Research (ICRAR), The University of Western Australia, 35 Stirling Hwy, Crawley, WA 6009, Australia}
\affiliation{CSIRO Space and Astronomy, PO Box 1130, Bentley, WA 6102, Australia}
\affiliation{International Centre for Radio Astronomy Research (ICRAR), Curtin University, Bentley, WA 6102, Australia}

\author{A.~T.~Deller}
\affiliation{Centre for Astrophysics and Supercomputing, Swinburne University of Technology, Hawthorn, VIC, 3122, Australia}

\author{T.~Dial}
\affiliation{Centre for Astrophysics and Supercomputing, Swinburne University of Technology, Hawthorn, VIC, 3122, Australia}

\author{K.~Gourdji}
\affiliation{Centre for Astrophysics and Supercomputing, Swinburne University of Technology, Hawthorn, VIC, 3122, Australia}

\author{S.~Simha}
\affiliation{University of California - Santa Cruz
1156 High St. Santa Cruz, CA, USA 95064}

\author{M. Caleb}
\affiliation{Sydney Institute for Astronomy (SIfA), School of Physics, The University of Sydney, Camperdown NSW 2006, Australia}

\author{L. Marnoch}
\affiliation{School of Mathematical and Physical Sciences, Macquarie University, NSW 2109, Australia}
\affiliation{Australia Telescope National Facility, CSIRO Space and Astronomy, P.O. Box 76, Epping, NSW 1710, Australia}
\affiliation{Astrophysics and Space Technologies Research Centre, Macquarie University, Sydney, NSW 2109, Australia}
\affiliation{ARC Centre of Excellence for All-Sky Astrophysics in 3 Dimensions (ASTRO 3D), Australia}

\author[0000-0002-7738-6875]{J. Xavier Prochaska}
\affil{University of California - Santa Cruz
1156 High St.
Santa Cruz, CA, USA 95064}
\affil{Kavli IPMU (WPI), UTIAS, The University of Tokyo, Kashiwa, Chiba 277-8583, Japan}
\affil{Division of Science, National Astronomical Observatory of Japan,
2-21-1 Osawa, Mitaka, Tokyo 181-8588, Japan}

\author[0000-0003-4501-8100]{S. D. Ryder}
\affiliation{School of Mathematical and Physical Sciences, Macquarie University, NSW 2109, Australia}
\affiliation{Astrophysics and Space Technologies Research Centre, Macquarie University, Sydney, NSW 2109, Australia}

\author{R.~M.~Shannon}
\affiliation{Centre for Astrophysics and Supercomputing, Swinburne University of Technology, Hawthorn, VIC, 3122, Australia}

\author{N. Tejos}
\affiliation{Instituto de F\'isica, Pontificia Universidad Cat\'olica de Valpara\'iso, Casilla 4059, Valpara\'iso, Chile}



\begin{abstract}

Identification and follow up observations of the host galaxies of fast radio bursts (FRBs) not only help us understand the environments in which the FRB progenitors reside, but also provide a unique way of probing the cosmological parameters using the dispersion measures of FRBs and distances to their origin. A fundamental requirement is an accurate distance measurement to the FRB host galaxy, but for some sources viewed through the Galactic plane, optical/NIR spectroscopic redshifts are extremely difficult to obtain due to dust extinction. Here we report the first radio-based spectroscopic redshift measurement for an FRB host galaxy, through detection of its neutral hydrogen (\HI) 21-cm emission using MeerKAT observations. We obtain an \HI-based redshift of $z = 0.0357\pm0.0001$ for the host galaxy of FRB\,20230718A, an apparently non-repeating FRB detected in the CRAFT survey and localized at a Galactic latitude of --0.367$^{\circ}$. Our observations also reveal that the FRB host galaxy is interacting with a nearby companion, which is evident from the detection of an \HI\ bridge connecting the two galaxies. A subsequent optical spectroscopic observation confirmed an FRB host galaxy redshift of 0.0359 $\pm$ 0.0004. This result demonstrates the value of \HI\ to obtain redshifts of FRBs at low Galactic latitudes and redshifts. Such nearby FRBs whose dispersion measures are dominated by the Milky Way can be used to characterise these components and thus better calibrate the remaining cosmological contribution to dispersion for more distant FRBs that provide a strong lever arm to examine the Macquart relation between cosmological DM and redshift.

\end{abstract}


\section{Introduction} \label{sec:intro}

Fast radio bursts (FRBs), first identified by \cite{Lorimer2007}, are highly energetic radio pulses occurring on timescales of milliseconds. In order to disentangle the various theories on FRB progenitors \cite[see review by][]{Cordes2019}, FRBs need to be localised to their host galaxies, and subsequently studied through follow-up observations. In addition, the `Macquart relation' between the dispersion measure (DM) and redshift \citep{Macquart2020} has been shown to reveal the previously `missing baryons' (hot ionised gas in the intergalactic medium), can resolve the Universe's large-scale structure \citep{RafieiRavandi2021,Lee2022}, and be used for cosmological studies \citep{James2022,Baptista2023}. Spectroscopic redshift measurements of the FRB hosts are essential for such studies. 

However, spectroscopic redshift information of the FRB
host galaxy candidates is not always straightforward to obtain. 
Indeed, the first precisely localized FRB \citep[FRB\,20121102A;][]{Chatterjee2017} 
occurred in an intrinsically
faint galaxy along a modestly extincted sightline through
the Galaxy. If placed at $z>1$ or behind another magnitude
of extinction, the identification of the host would become prohibitively expensive
for even the largest ground-based telescopes.
FRB hosts have been found at $z \sim 1$ \citep{Ryder2023}, while \cite{Marnoch2023} presented follow-up of FRB\,20210912A, which is still without an identified host despite deep observations with the Very Large Telescope (VLT) - even deeper optical time is required for these high-redshift FRBs through limited telescope resources. Another scenario is an FRB localised at low Galactic latitude, where dust extinction from the Galactic plane hampers optical follow-up. This is an issue for FRB surveys such as the Commensal Real-time ASKAP Fast Transients \cite[CRAFT;][]{Macquart2010,Bannister2017} survey, which collects data simultaneously with other scheduled observations, and so includes FRB searches during other major science surveys pointed toward the Galactic plane, with the Australian Square Kilometre Array Pathfinder telescope \cite[ASKAP;][]{Deboer2009,Hotan2021}. Other examples include the FAST telescope, where survey teams also conduct commensal FRB searches, such as with the Galactic Plane Pulsar Survey \cite[GPPS;][]{Han2021}, and the Meer(more) TRAnsients and Pulsars \cite[MeerTRAP;][]{Bezuidenhout2022}. Through GPPS observations \citet{Zhou2023} recently reported five FRBs all with Galactic latitude  $b<4^{\circ}$. 

Photometric redshifts are generally not accurate enough to be a reliable substitute for optical spectroscopic measurements in the Macquart relation. However, another alternative exists for FRBs at low Galactic latitudes: spectral line transitions in wavelength ranges that are less affected by dust extinction. For instance in the radio regime, the 21-cm (1420~MHz) line traces neutral hydrogen (\HI), the star-forming fuel in galaxies. The 1665-1667~MHz doublet for hydroxyl (OH) is typically associated with megamasers arising from starburst activity within the host galaxy. Such transitions are not without issues - for example, the strength of the 21-cm \HI\ transition decreases with redshift and is hard to detect at $z > 0.1$ due to both this effect and radio frequency interference (RFI) from artificial satellites affecting the $\sim$1150--1300~MHz regime ($0.09 < z_{\rm HI} < 0.23$), while the \HI\ radio beam size is often much larger (tens of arcseconds) than the FRB localisation. Nonetheless, it remains an avenue to obtain a spectroscopic redshift for FRB host galaxies when optical follow-up is difficult. \HI\ has been detected and studied in five FRB host galaxies to date, where four have shown signs of strong asymmetry in the \HI\ global spectrum and/or disturbed \HI\ intensity maps which were attributed to recent galaxy merger or interaction events \citep{Michalowski2021,Kaur2022,LeeWaddell2023}, and the fifth with less signs of interaction \citep{Glowacki2023}. Thus far, no non-detections of \HI\ have been reported in nearby FRB hosts, 
in line with the finding by \cite{Gordon2023} that FRB host galaxies tend to be in star-forming galaxies. Additionally, \cite{Hsu2023} presented asymmetric profiles of molecular gas (CO) in the host galaxy of FRB\,20180924B at $z = 0.3216$.

In this paper we present the first redshift for an FRB host galaxy whose redshift was measured originally from \HI\ in emission, and only later was confirmed from optical spectroscopy. In Section~\ref{sec:observations} we describe the FRB localisation and follow-up spectral-line observations in \HI. We present the \HI\ results in Section~\ref{sec:results}, and optical spectroscopic follow-up in Section~\ref{sec:optical}. In Section~\ref{sec:discussion} we briefly discuss the potential for future \HI\ follow-up studies, and summarise our findings in Section~\ref{sec:conclusions}. 


\section{Observations}\label{sec:observations}

\subsection{FRB detection and localisation}\label{sec:ASKAP}

\begin{figure}
\centering
\includegraphics[width=0.83\textwidth]{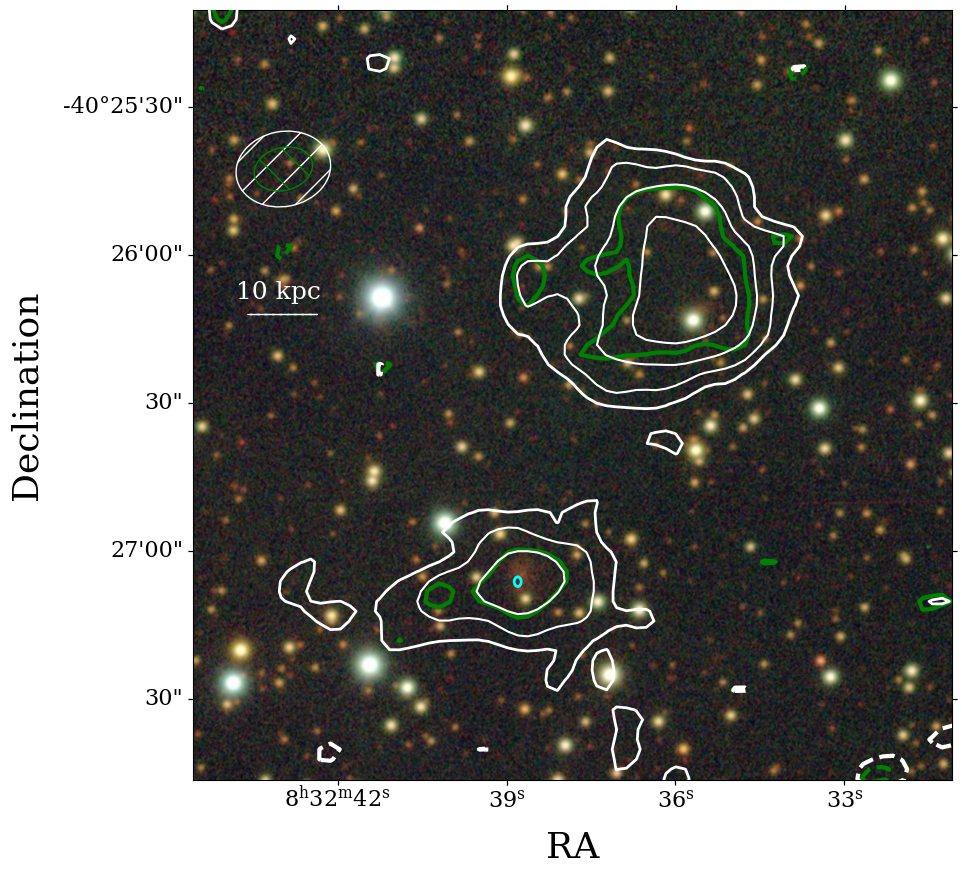}
\includegraphics[width=0.45\textwidth]{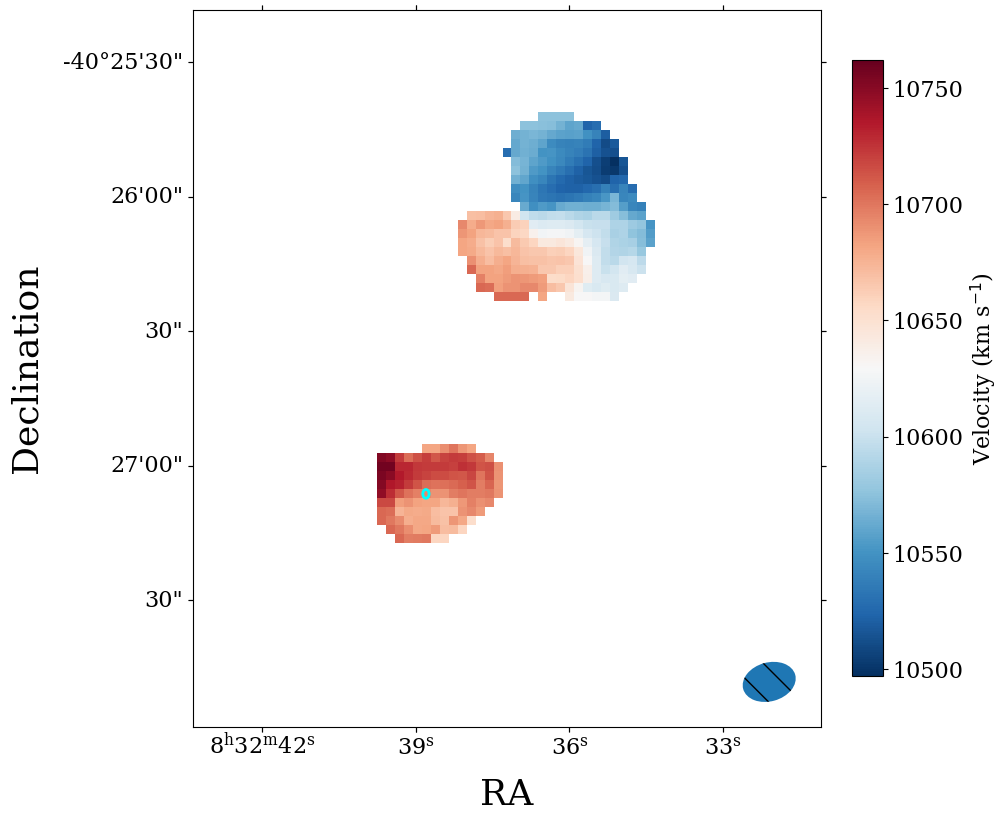}%
\includegraphics[width=0.45\textwidth]{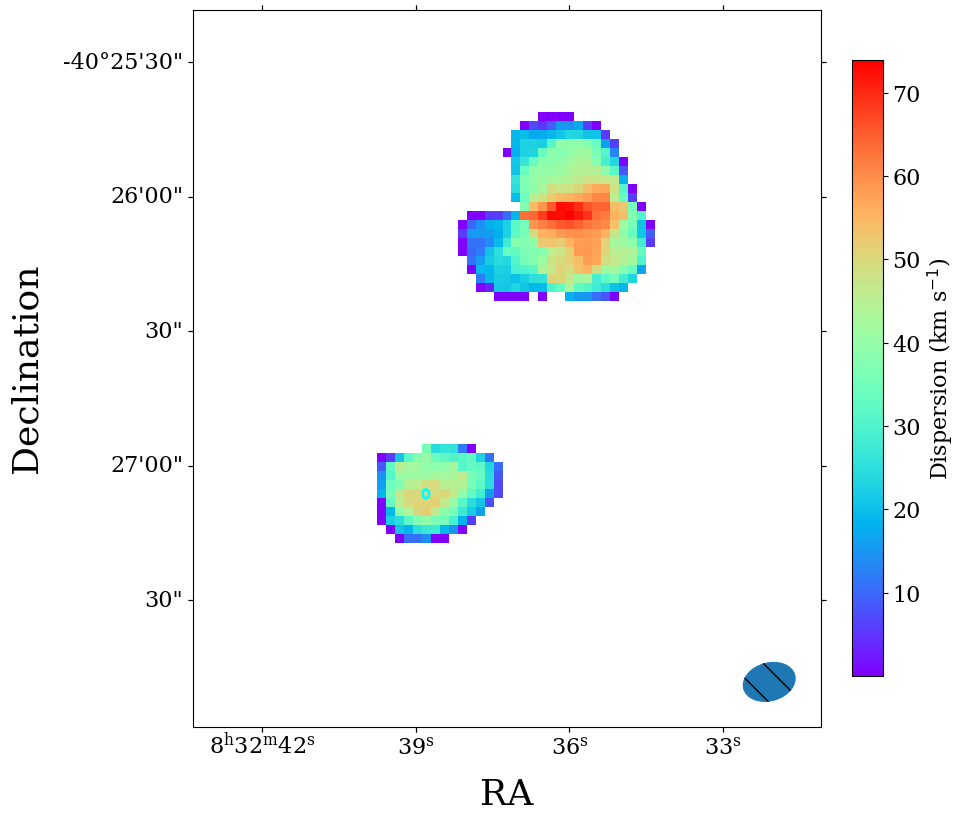}
\caption{Intensity, velocity, and dispersion (moments 0, 1, and 2) maps of the MeerKAT \HI\ DDT observations. Top: intensity map radio contours for the two higher spatial resolution cubes overlaid on DECam 3-colour image. For the lower spatial resolution of these two cubes (robust weighting of 1, radio beam of 19\farcs3 $\times$ 15\farcs2), contours are at column density multiples of 3 to the power of (1, 1.5, 2, 2.5, ...) of 2.4$\times$10$^{19}$~cm$^{-2}$, where the lowest contour is at 3$\sigma$ significance. Negative contours are given as dashed lines at the same levels. The 3$\sigma$ contour (green) is given for the 
higher resolution cube (robust weighting of 0.1, radio beam of 12\farcs0 $\times$ 8\farcs7), at 1.8$\times$10$^{20}$~cm$^{-2}$. The FRB localisation region is given by the cyan ellipse, enlarged by a factor of 5 for visibility, and is coincident with a peak in \HI\ emission and a faint, apparently red,  optical galaxy (noting extinction is a factor). Bottom panels: The highest spatial resolution spectral-line velocity and dispersion maps are displayed, produced and masked by SoFiA.}
\label{fig:momentmaps}
\end{figure}

During commensal observing with ASKAP, an FRB was successfully detected with a signal-to-noise ratio S/N of 10.9 on 18 August 2023 UT 07:02:08. Through the CRAFT Effortless Localisation and Enhanced Burst Inspection pipeline \cite[CELEBI;][]{Scott2023}, the FRB was localised to RA=08:32:38.804, Dec=--40:27:06.33 (J2000), with a 1$\sigma$ uncertainty ellipse of $0\farcs37$ in RA and $0\farcs39$ in Dec, with a position angle of --2.3$^{\circ}$. The S/N of the FRB in the post-processed image was 22.9. This FRB, henceforth FRB\,20230718A, was found to be coincident with the galaxy WISEA~J083238.73--402705.3, with no available redshift in the literature. The host galaxy is also seen in imaging for the DECam Plane Survey \cite[DECaPS2;][]{Saydjari2023} as an apparent faint red galaxy probably due to dust extinction (see Fig.~\ref{fig:momentmaps}). An analysis of the DECaPS2 coadded $r$-band imaging using Probabilistic Association of Transients to their Hosts \cite[PATH;][]{Aggarwal2021} under standard priors, correcting for Galactic extinction using the \citet{Schlafly2011} dust maps and the \citet{FM07} reddening law, yields a PATH posterior probability of 94\% that this object is the host galaxy. Further analysis of the burst profile of FRB\,20230718 and the polarisation properties will be presented in Scott et al., (in prep.). 

The measured DM for FRB\,20230718A from CELEBI is 476.6~pc\,cm$^{-3}$, which ordinarily would imply via the Macquart relation a host galaxy redshift $z > 0.3$. However, given the Galactic latitude of the host galaxy is --0.367$^{\circ}$, the contribution of the Milky Way at this position is found to range from 393 or 421~pc\,cm$^{-3}$ from the Baror/Prochaska\footnote{\url{https://github.com/FRBs/ne2001}} or PyGEDM \cite[Python Galactic electron density model;][]{Price2021} implementation for the NE2001 model  \citep{Cordes2002,Cordes2003}, to as high as 450~pc\,cm$^{-3}$ from the YMW16 model \citep{Yao2017}. Hence the extragalactic DM would be 26~$<$~DM$_{\rm EG}$~$<$~83~pc\,cm$^{-3}$, consistent with a host galaxy at $z < 0.1$. We note that the cosmic DM, DM$_{\rm cosmic}$ = DM$_{\rm EG}$ - DM$_{\rm host}$, would be smaller than the extragalactic DM$_{\rm EG}$ when requiring the host galaxy DM $> 0$~pc\,cm$^{-3}$. High Galactic extinction \cite[$>3$~mag in most bands;][]{Schlafly2011} is likely the reason why no optical spectroscopic redshift was available for the galaxy WISEA~J083238.73--402705.3, and motivated an alternative approach for obtaining a spectroscopic redshift. 

\subsection{Searching for the host galaxy of FRB\,20230718A in H{\textsc i}-21 cm emission}

We began by searching for archival \HI\ datasets. No detection was seen in the \HI\ Parkes All Sky Survey \cite[HIPASS;][]{Barnes2001}. We note that the 3$\sigma$ $M_{\rm HI}$ sensitivity in HIPASS is stated to be 10$^{6}$\,$d_{\rm Mpc}^{2}$\,M$_{\odot}$, which corresponds to 
$\sim$2.9$\times$10$^{10}$~M$_{\odot}$ at the HIPASS distance limit of 170~Mpc (table 1 of \cite{Barnes2001}).
A search of the MeerKAT archive\footnote{\url{https://apps.sarao.ac.za/katpaws/archive-search}} hosted by the South African Radio Astronomical Observatory (SARAO) indicated various pointings in L-band which contained the FRB position by the `Legacy Survey of the Galactic Plane', project ID SSV-20180721-FC-01. Three observations with capture block IDs of 1574801265, 1574911560, and 1574542866 were identified with the FRB position within $0.5^{\circ}$ of the field centre of each pointing; two of the observations had the same pointing for a combined 1~hr of on-source time. A preliminary inspection of the data revealed weak \HI\ emission at the FRB position. This motivated a deeper follow-up observation in L-band through Director Discretionary Time (DDT). Observation 1697669539 was centred on the FRB localisation, and carried out on 18--19 October 2023 with MeerKAT as part of proposal ID DDT-20231017-MG-01 (see Table~\ref{tab:obsdetails}). The observation was taken in L-band (856~MHz bandwidth centred at 1284~MHz) at 32K spectral resolution (26.123~kHz wide channels, corresponding to 5.7~km\,s$^{-1}$ channels at the detected \HI\ emission frequency). The SARAO SDP continuum image quality report gave a root mean square (RMS) noise of 11~$\mu$Jy. We simultaneously searched for repeats bursts in realtime using the MeerTRAP single pulse search pipeline as described in \cite{chr+22, rbc+22}. No repeat bursts were detected above a fluence threshold of 0.09~Jy ms for a 1~ms wide burst around the DM of the FRB.

\begin{table}
\small
\caption{Details of the MeerKAT observation and spectral-line cubes for proposal ID DDT-20231017-MG-01.}
\centering
\begin{tabular}{ll}
\hline\hline
SBID & 1697669539\\
Phase centre & 08:32:38.804, --40:27:06.33\\
Bandpass calibrator & J1939-6342\\
Gain calibrator & J1744-5144\\
Channel width & 26.123~kHz (later rebinned by factor of 2)\\
On-source integration time & 4 hr\\
Highest spatial res cube & UV range of 0--40k$\lambda$, radio beam of 12.0"$\times$8.7", PA of 105$^\circ$, 2" pixels, robust weighting of 0.1 \\
 & 2$\sigma$ $N_{\rm HI}$ sensitivity of 1.2$\times$10$^{20}$~cm$^{-2}$ in intensity map \\
High spatial res cube & UV range of 0--40k$\lambda$, radio beam of 19.3"$\times$15.2", PA of 100$^\circ$, 2" pixels, robust weighting of 1 \\
 & 2$\sigma$ $N_{\rm HI}$ sensitivity of 4.7$\times$10$^{19}$~cm$^{-2}$ in intensity map \\
Low spatial res cube & UV range of 0--17.5k$\lambda$, radio beam of 33.2"$\times$18.9", PA of 90$^\circ$, 5" pixels, robust weighting of 1 \\
 & 2$\sigma$ $N_{\rm HI}$ sensitivity of 1.2$\times$10$^{19}$~cm$^{-2}$ in intensity map \\
Lowest spatial res cube & UV range of 0--10k$\lambda$, radio beam of 38.6"$\times$24.9", PA of 93$^\circ$, 5" pixels, robust weighting of 1 \\
 & 2$\sigma$ $N_{\rm HI}$ sensitivity of 5.7$\times$10$^{18}$~cm$^{-2}$ in intensity map \\
\hline
\end{tabular}
\label{tab:obsdetails}
\end{table}

\begin{figure}
\centering
\includegraphics[width=0.83\textwidth]{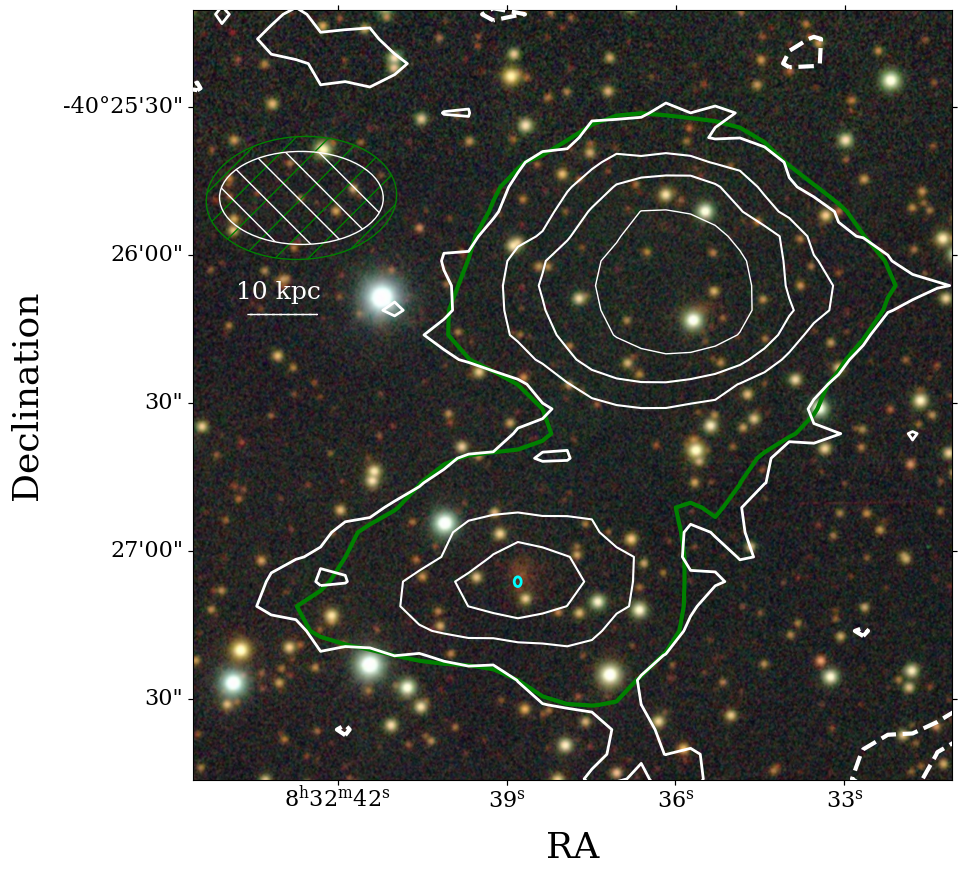}
\includegraphics[width=0.45\textwidth]{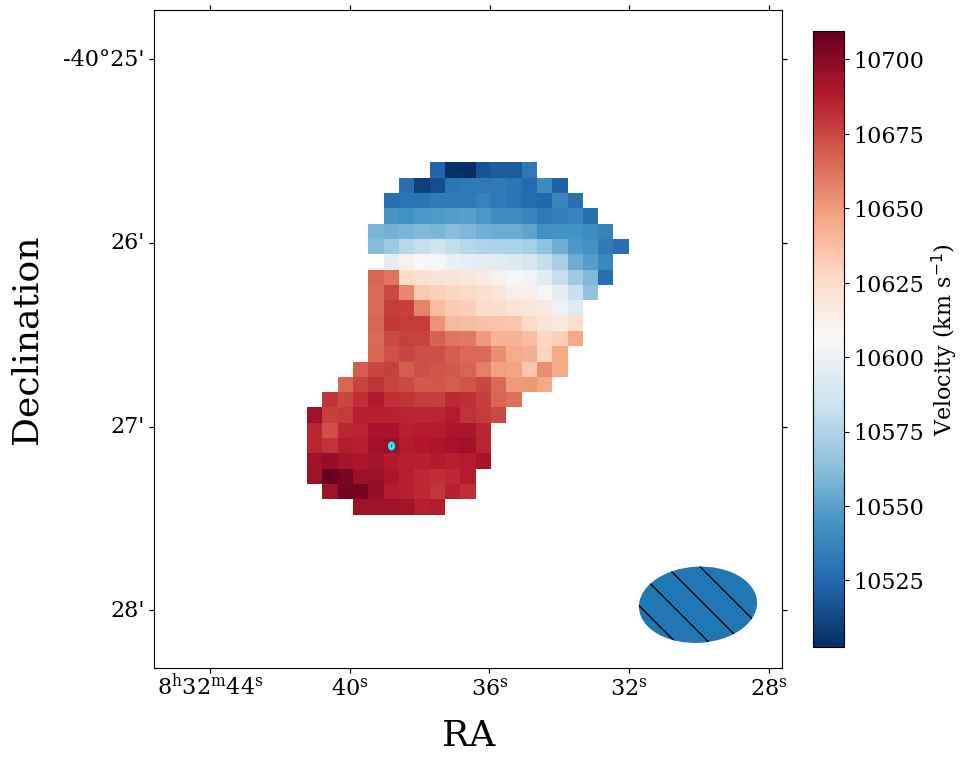}%
\includegraphics[width=0.45\textwidth]{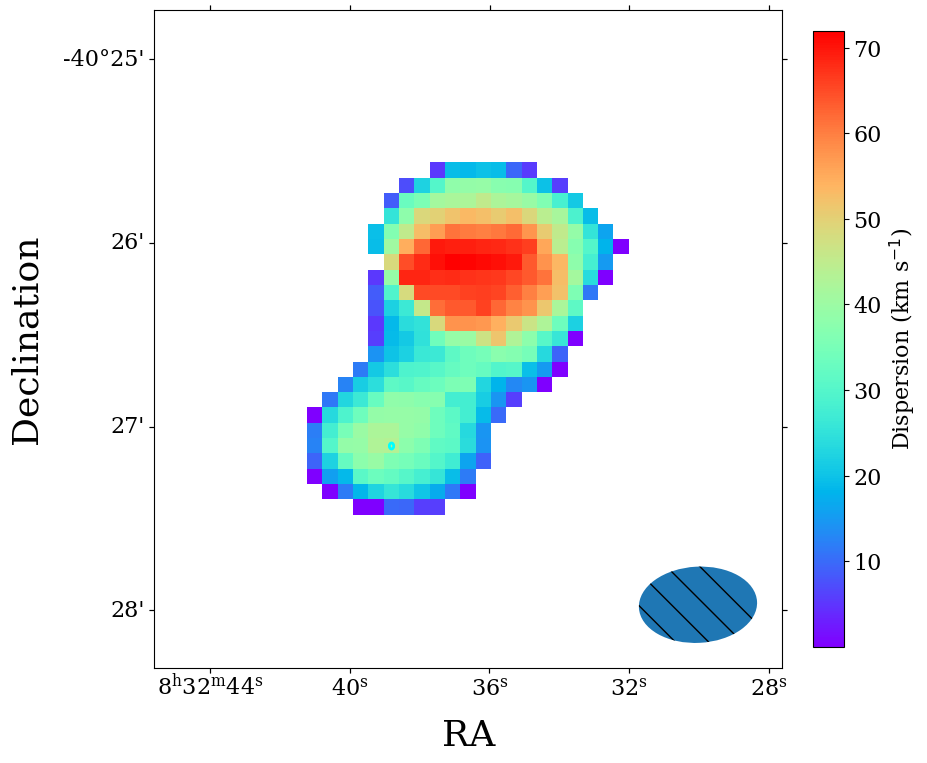}
\caption{Moment maps as in Fig.~\ref{fig:momentmaps}, for the two lower resolution cubes. Contours are at column density multiples of 3 to the power of (1, 1.5, 2, 2.5, ...) of 8.5$\times$10$^{18}$~cm$^{-2}$ for the `low' (17.5\,k$\lambda$, radio beam of 33\farcs2$\times$18\farcs9) resolution cube here (white), where the lowest contour is at 3$\sigma$ significance, and negative contours are given as dashed lines. The 3$\sigma$ contour (green) is given for the intensity map for the lowest resolution cube (radio beam of 38\farcs6 $\times$ 24\farcs9) at 8.6$\times$10$^{18}$~cm$^{-2}$. The velocity and dispersion maps are produced from the lowest resolution cube, again through SoFiA.}
\label{fig:momentmap_lowrescube}
\end{figure}

The raw data were transferred to the {\tt ilifu} supercomputing cloud system and reduced there. Bandpass, flux, and phase calibration, along with self-calibrated continuum imaging, was performed using the {\sc processMeerKAT} pipeline\footnote{\url{https://idia-pipelines.github.io/docs/processMeerKAT}}, which is written in Python, uses a purpose-built CASA \citep{McMullin2007} Singularity container, and employs MPICASA (a parallelised form of CASA). Data was at this stage rebinned by a factor of 2 (i.e. to 16K mode, 52.246~kHz-wide channels). Model continuum visibility data were subtracted from the corrected visibility data using the CASA task \textit{uvsub}. A first-order polynomial fit to the continuum was then calculated and subtracted using the CASA task \textit{uvcontsub} for all channels to remove residual continuum emission from the spectral line data, with known \HI\ emission channels excluded from the continuum fit. Finally, four spectral line cubes were created using CASA task \textit{tclean} with robust~=~1 (0.1 for the highest resolution cube) and cleaning to 1.5$\times$RMS (root mean square) levels, and variable uv-distances for the uvrange parameter and pixel sizes with no uvtaper used (see Table~1). Using a UV taper did not change the resulting cubes significantly. The RMS per 52.246~kHz channel was fairly stable across the middle of the band, with a per-channel noise of 0.18 and 0.16~mJy\,beam$^{-1}$ at the \HI\ emission for the FRB host galaxy for the low and high resolution cubes. All channels were convolved to a common synthesized beam for each spectral-line cube, and katbeam\footnote{\url{https://github.com/ska-sa/katbeam}} used for primary beam correction. The Source Finding Application 2 \cite[SoFiA 2;][]{Serra2015,Westmeier2021} program was run on each cube. 

\section{Results}

\subsection{HI properties}\label{sec:results}

\begin{table}[t]
\small
\caption{Measured properties of the host galaxy FRB\,20230718A and its neighbor galaxy in order of \HI\ flux, \HI\ mass, \HI\ distance, \HI\ 21cm frequency Freq$_{\rm HI}$, \HI\ position, and radio continuum flux, as measured from the lowest spatial resolution cube (except for $Freq_{\rm HI}$ and \HI\ position, where measurements from the distinct SoFiA detections in the second highest resolution cube are used instead).}
\centering
\begin{tabular}{llll}
\hline\hline
Quantity & FRB host & {\bf Neighbor} & Combined system \\
 \hline
$S_{\rm HI}$ & 0.137\,$\pm$\,0.010~Jy~km\,s$^{-1}$ & 0.363\,$\pm$\,0.014~Jy~km\,s$^{-1}$ & 0.557\,$\pm$\,0.021~Jy~km\,s$^{-1}$\\ 
$M_{\rm HI}$ & 8.28\,$\pm\,$0.32$\times$10$^{8}$~M$_{\odot}$ & 2.15\,$\pm$\,0.10$\times$10$^{9}$~M$_{\odot}$ & 3.33\,$\pm$\,0.13$\times$10$^{9}$~M$_{\odot}$\\
$D_{\rm HI}$ & 162.9~Mpc & 161.2~Mpc & 162.1~Mpc\\
Freq$_{\rm HI}$ & 1371.49~MHz & 1371.96~MHz & --\\
\HI\ Position & 08:32:38.642, --40:27:05.43 & 08:32:35.992, --40:26:04.56 & --\\
$S_{\rm cont}$ & 0.97~mJy & Host unclear & --\\
\hline
\hline
\end{tabular}
\label{tab:galdetails}
\end{table}

In Figures~\ref{fig:momentmaps} and \ref{fig:momentmap_lowrescube} we present \HI\ intensity maps, overlaid on a DECaPS2 3-colour image ($g,r,z$ bands), accompanied by the velocity and dispersion maps for the high and low resolution \HI\ cubes respectively. \HI\ emission was seen for both the FRB host galaxy with the peak overlapping with the FRB localisation, and a neighboring \HI-rich galaxy. The latter is unable to be attributed to an optical counterpart due to a few candidates appearing within the highest \HI\ contour level and beam size; this source is henceforth referred to as the `neighbor' or `neighboring galaxy'
. Using the SoFiA measured \HI\ position for the neighboring galaxy, the projected separation distance between it and the FRB localisation is 55~kpc (69\farcs6). We also detect a \HI\ bridge connecting the two, at $>8\sigma$ sensitivity in the lowest resolution (38\farcs6$\times$24\farcs9) masked cube ($>3\sigma$ in the second lowest resolution (33\farcs2$\times$18\farcs9) unmasked cube), or to a \HI\ column density of least 4$\times$10$^{19}$~cm$^{-2}$. At the highest spectral resolution (12\farcs0$\times$8\farcs7) we see some evidence of rotation in the FRB host (lower-left panel of Fig.~\ref{fig:momentmaps}).

\begin{figure*}
\centering
\vspace{-0.2cm}
\includegraphics[width=0.85\textwidth]{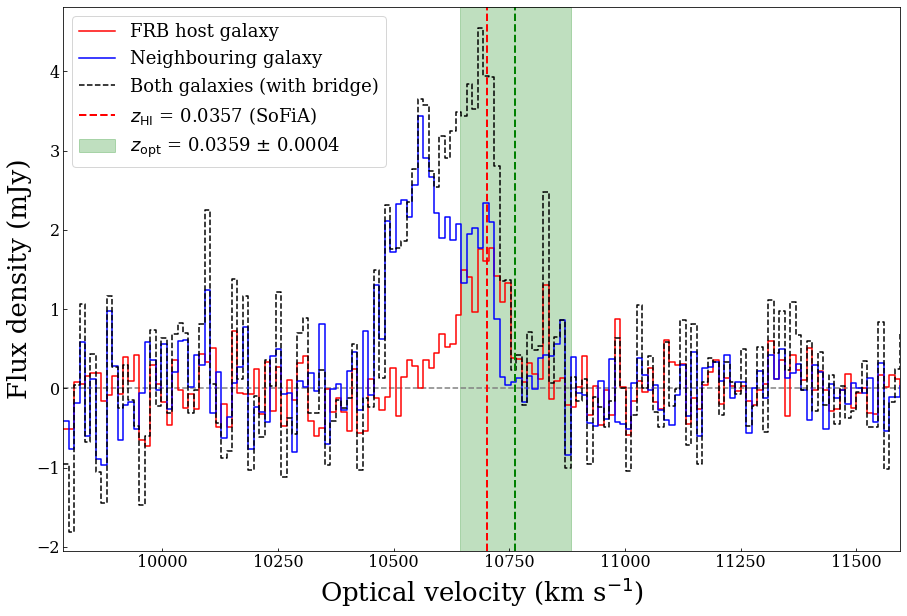}
\includegraphics[width=\textwidth]{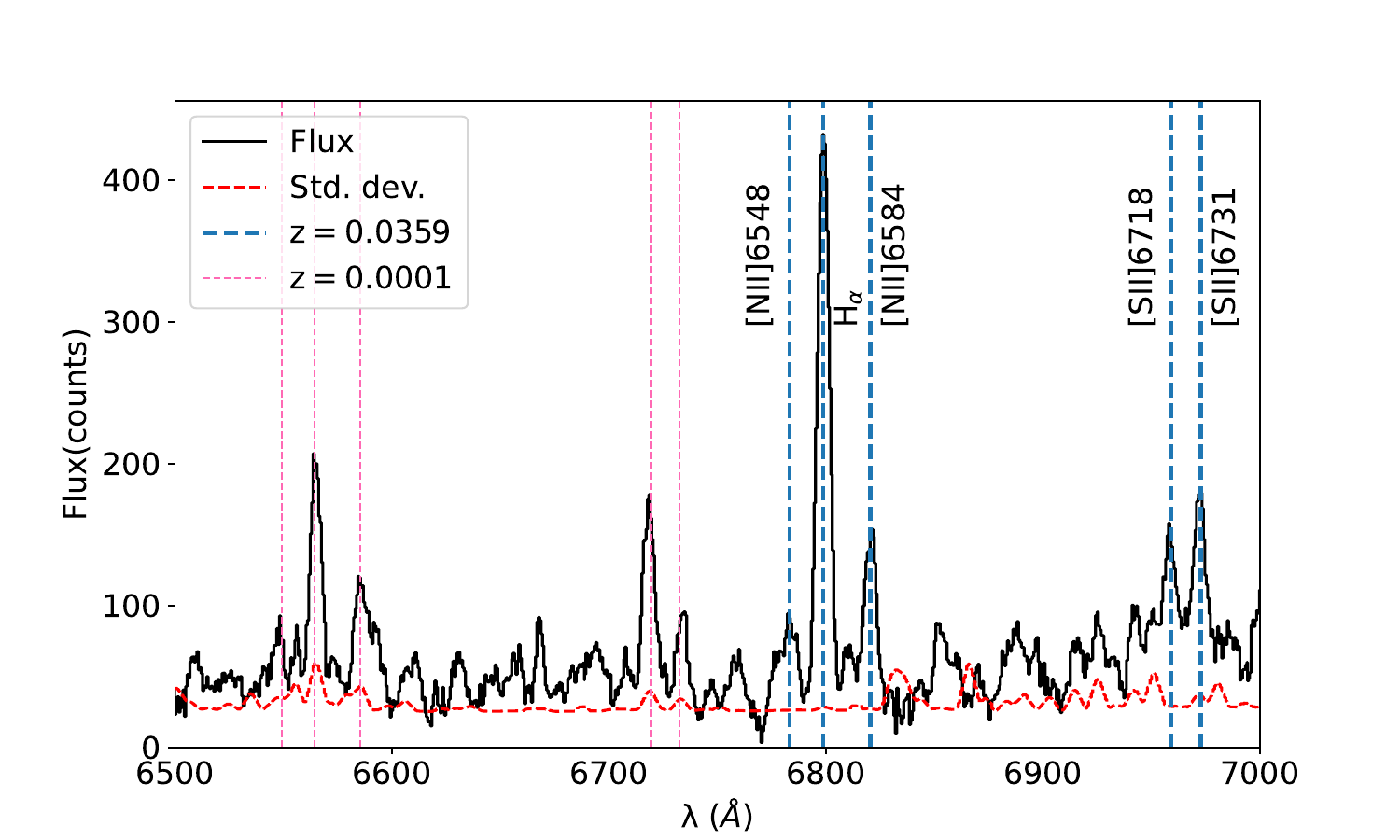}
\caption{Top: \HI\ spectra for the FRB host galaxy (red), the neighboring galaxy (blue), and the total combined system including the \HI\ bridge (black). 
The green shaded region indicates the optical spectroscopic redshift and error obtained in the follow-up optical spectroscopy. Bottom: DEIMOS spectrum of the host galaxy of FRB20230718A. The locations of the detected emission lines in the spectrum (black) are marked with blue dashed lines. The standard deviation per wavelength bin is shown by the dashed red line. The same emission lines ($H\alpha$, [N\,{\sc ii}] and [S\,{\sc ii}]) from the Milky Way at $z=0.0001$ are also visible (thinner pink lines) but the names of the transitions are not marked to avoid clutter.}
\label{fig:spectra}
\end{figure*}

In Figure~\ref{fig:spectra} we present \HI\ spectra for the putative FRB host galaxy, the neighboring galaxy, and the combined region including the \HI\ bridge, from the lowest resolution unmasked cube. 
The presence of the \HI\ bridge, as well as a somewhat lopsided \HI\ spectrum with additional flux on the lower velocity side, shows that the FRB host galaxy distribution is interacting with its neighbor. The neighboring galaxy appears to be regularly rotating with a lopsided double-horned spectral-line profile. 
We do not detect any other \HI\ sources along the line of sight to the FRB host galaxy, although we do detect several other \HI\ galaxies in emission at a similar redshift elsewhere in the cube (0.4--1.9~Mpc projected distances from the FRB host), indicating this FRB host galaxy resides in a \HI-rich galaxy group.

We summarise the following properties calculated from the \HI\ data and ancillary data in Table~\ref{tab:galdetails}, using the \HI\ redshift of $z~=~0.0357\pm0.0001$ \cite[corresponding to a distance of 162.8~Mpc assuming Planck cosmology and $H_{\rm 0}$ = 67.7 km\,s$^{-1}$\,Mpc$^{-1}$;][]{Planck2016}. 
We note that it is difficult to distinguish between the \HI\ associated with the FRB host galaxy and the bridge connecting it to its neighbor. The companion is more \HI\ massive than the FRB host by 2.7 times (or $\sim$65\% of the combined \HI\ mass). 

An unresolved radio continuum source with a flux density of 0.97~mJy (from 1304--1420 MHz) is associated with the FRB and \HI\ emission at $\sim5\sigma$ significance. Assuming this radio continuum is entirely due to star formation, we use the method described in \cite{Grundy2023} (equations 9 and 10), which applied the methodology of \cite{Molnar2021} to galaxies detected in radio continuum at 1.3675~GHz to estimate the total global star formation rate (SFR). For the FRB host we find the SFR to be 1.69$\substack{+0.49 \\ -0.50}$~M$_{\odot}$\,yr$^{-1}$. We note that we are using a radio flux density measured between 1304--1420~MHz rather than at 1400~MHz as in \cite{Molnar2021}, but \cite{Grundy2023} found that their unresolved 1.3675~GHz radio continuum fluxes were consistent with 1.4~GHz measurements within the scatter. The WISE colours of $(W1-W2) = 0.286 \pm 0.075$ and $(W2-W3) = 4.107 \pm 0.092$ suggest that the host of FRB\,20230718A is unlikely to harbor a highly efficient accreting AGN as it fits among the normal star-forming spiral region of the Wide-field Infrared Survey Explorer (WISE) colour-colour diagram \cite[see fig. 10 of][]{Wright2010}. The red $(W2-W3)$ colour is in part attributed to Galactic extinction. 


\subsection{Optical spectroscopic confirmation}\label{sec:optical}

On Dec 14 2023 (UTC) we obtained an optical spectrum with the DEep Imaging Multi-Object Spectrograph \cite[DEIMOS;][]{Faber2003} on the Keck telescope.  To ensure a fast and accurate acquisition of the host, we designed a slitmask with 5 square slits for alignment stars and one slit centered on the coordinates of the host candidate
from the Pan-STARRS optical imaging database. 
We used the 600ZD grating set to a central wavelength of 6500\AA\ combined with the GG455 order-blocking filter to provide a nominal,
total wavelength coverage of 4000-8700\AA.
Owing to a detector failure on this night,
the raw frames acquired with the ``Single:B'' readout mode were limited to $\lambda = 6000-8700$\AA.
Three 900s exposures were obtained for a total of 2700s. In addition to the science exposures, we obtained an arc frame for wavelength calibration and a set of flat-frame images for slit edge identification and flatfielding.

The spectrum was reduced using the PypeIt\footnote{https://github.com/pypeit/PypeIt} package \citep[v. 1.13;][]{pypeit:joss_pub, pypeit:zenodo}. PypeIt automatically subtracts the bias levels, traces the individual slit edges, performs flexure and flat-field corrections, and masks cosmic rays. After the basic processing, PypeIt detects the object continuum, performs sky subtraction in each slit individually,
extracts the spectrum, and applies the 
wavelength calibration. 
We did not apply a flux calibration as our goal was to obtain the redshift alone. The individual exposures were reduced using the default PypeIt parameters, and the extracted 1D spectra were coadded to produce a final spectrum. We used the MARZ redshifting software \citep{MARZ}. MARZ cross-correlated a template spectrum against the host spectrum and determined that the optimal redshift is 0.03591. While the software does not produce an error estimate for the spectroscopic redshift, we used the location of the half-maximum of the spectral lines relative to the peak as a measure of the redshift bound, i.e., $\delta z = 0.0004$. 

The bottom panel of Fig.~\ref{fig:spectra} shows our final DEIMOS spectrum. At $z=0.0359 \pm 0.0004$, $H\alpha$, the [N\,{\sc ii}] doublet (6548\,\AA, 6584\,\AA), and the [S\,{\sc ii}] doublet (6716\,\AA, 6731\,\AA) were clearly detected, thus confirming the redshift we measured in \HI\ from SoFiA. We also detected a 
faint [O\,{\sc iii}] 5007\,\AA\ line at the same redshift. In addition to the host galaxy emission, we also detect $H\alpha$, [N\,{\sc ii}] and [S\,{\sc ii}] emission at $z\sim0.0001$, arising from the Milky Way. Therefore, we confidently rule out the possibility that the \HI\ emission detected here is merely a chance coincidence with the FRB host galaxy (see also Section~\ref{sec:HIzdiscussion}).


\section{Discussion}\label{sec:discussion}

\subsection{HI environment}

The FRB host galaxy is not isolated, given the clear detection of a \HI\ bridge between it and the neighbor galaxy. Therefore, this would be the fifth FRB host galaxy with a sign of galaxy interaction within the \HI\ observations to date out of six published so far, akin to \cite{Michalowski2021,Kaur2022,LeeWaddell2023}. 
Given the presence of the \HI\ bridge and the active star formation seen in the FRB host galaxy (assuming all of the radio continuum is attributed to star formation), it is possible that the FRB host galaxy is accumulating gas from the neighbor. This would align with the proposed `fast FRB channel model' discussed in \cite{Michalowski2021}, where a galaxy interaction event triggers star formation, leading to the birth of massive and relatively short-lived stars that result in the creation of neutron stars or magnetars, which are popular FRB progenitor candidates. 

Additionally, at least seven other \HI\ emission detections within the same MeerKAT observation have been made within a 5~MHz block containing the FRB host galaxy \HI\ emission, identified from visual inspection and confirmed through SoFiA. This FRB host hence resides in a \HI-rich galaxy group. In cross-matching four optical galaxy group catalogues with the Arecibo Legacy Fast ALFA Survey \cite[ALFALFA;][]{Haynes2018}, \cite{Jones2020} found that ALFALFA contained a far greater number of field galaxies rather than group members, with no region of the environment parameter space explored with group galaxies as the dominant population. 


\subsection{FRB host galaxy redshifts}\label{sec:HIzdiscussion}

This is the first demonstration of using a radio spectral transition to measure the redshift of an FRB host galaxy. Currently, one limitation arises from the diminishing intensity of the \HI\ 21-cm transition with redshift, where only the most \HI\ massive galaxies would be visible with even a full night of observation with MeerKAT above $z > 0.23$. RFI limits \HI\ studies between $0.09 < z < 0.23$. While OH megamaser emission seen through the 1665--1667~MHz doublet spans a different redshift space not affected by RFI (e.g. RFI affects the $0.27 < z < 0.44$ redshift space for this transition), such detections are less plentiful compared to \HI\ as they are typically only associated with starburst galaxies. Little more than 100 OH megamasers exist in the current literature, although the advent of SKA pathfinders are poised to improve on this space \citep{Roberts2021}.

\HI\ searches in absorption avoid the issue of a fall-off in intensity of the transition with redshift seen for emission-line searches, but would require either the FRB host to be a sufficiently bright radio galaxy to enable such a search, or a search within the FRB signal itself. The latter is particularly interesting as it would yield information on both the cold neutral hydrogen and the hot ionised gas within the FRB sightline \cite[see][]{Fender2015}. No \HI\ absorption was seen from a preliminary analysis of the FRB\,20230718A burst profile, although we note the measured S/N of 22.9 of the FRB burst meant we were not sensitive to the column density of \HI\ we detected here, when assuming a spin temperature of 100~K and high optical depth. Such a detection of \HI\ absorption in an FRB signal would only be possible in the brightest FRBs and require they be detected in the L-band. A stack of many ($\sim$100) FRB signals at low Galactic latitudes may provide an avenue for an \HI\ absorption detection\footnote{See poster by Om Gupta at FRB2023: \url{https://drive.google.com/file/d/13VdW-5XmHbVCRqf6LQsGKRq1ANaA_RgT/view}}. However, this stacking approach requires knowing the FRB redshift and so cannot be used to measure the redshifts of individual FRBs.

We briefly consider the possibility of the observed \HI\ emission aligning with the FRB sky localisation by chance rather than being truly associated with the FRB host galaxy, had we not obtained a more recent optical spectroscopic measurement. For this, we take the number of \HI\ sources from the ALFALFA survey (31,502), conducted over 7,000 square degrees of sky, to arrive at a source density of $\sim$4.5 per square degree up to $z \sim 0.06$. While the \HI\ mass sensitivity, and the expected corresponding size of \HI\ in each galaxy ALFALFA is sensitive to, varies as a function of distance; and the presence (or absence) of any large-scale structure will greatly vary the expected number of \HI\ sources, we expect 0.00125 \HI\ sources in any square arcminute of sky based on ALFALFA. Therefore we consider the chance of detecting \HI\ from a separate galaxy to the FRB host to be negligible. We searched up to $z_{\rm HI} \sim 0.09$.

We briefly consider different applications from using the 21-cm \HI\ transition for determining redshifts for FRB host galaxies. Potentially 10s to 100s of FRBs a year will be found through the Galactic plane. The host galaxies of these FRBs are both useful for specific scientific goals, and difficult to obtain via optical spectroscopy. The first application of low Galactic latitude FRBs, specifically those originating in nearby galaxies, is to assist in characterising the Milky Way disk and halo DM models. Nearby FRBs have the twin benefit that the hosts can most easily be identified in \HI\ emission, while the uncertainty in the minimum non-Milky Way contribution can be very small for low-DM bursts. Currently, halo models are uncertain - the DM contributed by Milky Way halo DM$_{\rm halo}$ has been estimated to range 10--80~pc\,cm$^{-3}$ \citep{Prochaska2019,Keating2020}, and is often simply assumed to be 50~pc\,cm$^{-3}$ \cite[e.g.][]{James2022a}. For FRB\,20230718A, with DM$_{\rm EG}$ ranging between $\sim$26--83~pc\,cm$^{-3}$, and assuming 10~pc\,cm$^{-3}$ per 0.01 in redshift (as done in \cite{James2022a} for the Macquart relation), DM$_{\rm cosmic}$ would be $\sim$35.7~pc\,cm$^{-3}$, placing an upper limit of both DM$_{\rm host}$ and DM$_{\rm halo}$ combined of $\sim$47.3~pc\,cm$^{-3}$. Even assuming the lower end for DM$_{\rm halo}$ would leave a fairly modest amount of DM contributed by the FRB host galaxy.

Another application is using the distances derived from \HI\ measurements for FRB hosts to help determine the distances of scattering screens to the FRB source  \citep{Sammons2023}, given that Galactic scintillation can be assumed to be a dominating effect \cite[e.g.][]{Chawla2022}. \cite{Sammons2023} determined constraints for the distances of scattering screens of a few CRAFT FRBs, through analysis of high-time resolution datasets available through CELEBI \citep{Scott2023}. Redshift measurements for such FRB host galaxies through the \HI\ line hence can help further constrain or rule out such models on the scattering screen distances (see e.g. equation 3 of \cite{Sammons2023}). This in turn can aid model comparison for FRB progenitors and their immediate environments, alongside disentangling estimations of both DM$_{\rm host}$ and DM$_{\rm cosmic}$ \citep{Ocker2022}.

Nearby FRBs away from the Galactic plane could also be localised through the \HI\ 21-cm transition. The number of FRBs localised will improve from an expected increase in FRB detection rates, such as for CRAFT through its CRACO (CRAFT COherent) upgrade to the real-time detection system. While most FRBs detected by CHIME are not well constrained in their position to enable redshift searches, new localisations for one-off and repeating bursts have been reported \citep{Bhardwaj2023,Ibik2023}, while the CHIME Outriggers project will detect and localize FRBs with 50 mas precision \citep{MenaParra}. The 110-antenna Deep Synoptic Array (DSA-110) telescope also recently shared localisations for 11 FRBs \citep{Law2023}. A lower localisation precision is sufficient to obtain a host galaxy for nearby FRBs, compared to higher redshift FRBs. Previously conducted large \HI\ surveys such as HIPASS and ALFALFA will be searchable against any such FRB localisations, but we note the coarse angular resolution of such surveys ($\sim$3 arcminutes for ALFALFA, $\sim$15 arcminutes for HIPASS) is a limiting factor in conclusively identifying FRB host galaxies. Of more benefit will be the Widefield ASKAP L-band Legacy All-sky Blind surveY \cite[WALLABY;][]{Koribalski2020}. WALLABY has a spatial resolution of $30\arcsec$, and is currently ongoing and resulted in the first commensal detection of an FRB and the \HI\ in its host galaxy \citep{Glowacki2023}. WALLABY offers an improved angular resolution and depth ($z < 0.09$) to ALFALFA and HIPASS, and so will enable association of nearby FRBs to hosts without the need for dedicated optical follow-up. Probabilistic associations of nearby FRBs through PATH could hence be applied to \HI\ data, rather than optical datasets, but an estimate of the probability of \HI\ 21-cm emission interlopers at this resolution would be a necessary part of any such PATH analysis. 

\section{Conclusions}\label{sec:conclusions}

We have presented \HI\ 21cm follow-up observations from the MeerKAT radio telescope of FRB\,20230718A, which was localised to sub-arcsecond precision to a faint dust-extincted galaxy at low Galactic latitude. \HI\ in emission was detected for the host galaxy, as well as a nearby companion galaxy with no clear optical host in the available optical photometry. An \HI\ bridge is seen between the two galaxies, indicating that the host of FRB\,20230718A is interacting and potentially undergoing enhanced star formation, which could have led to a progenitor formed from a massive star, as seen in four other FRB host galaxies detected in \HI. 

From this observation we obtained a redshift of 0.0357, the first redshift measurement for an FRB host galaxy not derived first from optical observations. We have since confirmed that the \HI\ emission seen here is associated with the FRB host through optical spectroscopic follow-up. This demonstrates the ability to use the \HI\ 21-cm transition to follow up FRB host galaxies for which it may be difficult to obtain optical spectroscopic measurements, such as for other FRB hosts behind the Galactic plane. Redshift information obtained through \HI\ provides an avenue to including such localised FRBs in studies such as constraining the DM contribution from the Milky Way and FRB host galaxy, FRB progenitor models, and cosmological studies. 

\section{Acknowledgements}

We thank the anonymous referee for the useful feedback provided which has improved the paper. We thank Ben Stappers for assistance with the MeerTRAP single pulse search pipeline. MG is supported by the Australian Government through the Australian Research Council's Discovery Projects funding scheme (DP210102103). RMS and ATD acknowledge support through Australian Research Council Future Fellowship FT190100155 and Discovery Project DP220102305. MC acknowledges support of an ARC Discovery Early Career Research Award DE220100819 funded by the Australian Government and the ARC Centre of Excellence for All Sky Astrophysics in 3 Dimensions (ASTRO 3D), through project number CE170100013. KG acknowledges support through Australian Research Council Discovery Project DP200102243. LM acknowledges the receipt of an MQ-RES scholarship from Macquarie University. Authors JXP and NT, as members of the Fast and Fortunate for FRB Follow-up team, acknowledge support from NSF grants AST-1911140, AST-1910471 and AST-2206490.

This scientific work uses data obtained from Inyarrimanha Ilgari Bundara, the CSIRO Murchison Radio-astronomy Observatory. We acknowledge the Wajarri Yamaji People as the Traditional Owners and native title holders of the Observatory site. CSIRO’s ASKAP radio telescope is part of the Australia Telescope National Facility (https://ror.org/05qajvd42). Operation of ASKAP is funded by the Australian Government with support from the National Collaborative Research Infrastructure Strategy. ASKAP uses the resources of the Pawsey Supercomputing Research Centre. Establishment of ASKAP, Inyarrimanha Ilgari Bundara, the CSIRO Murchison Radio-astronomy Observatory and the Pawsey Supercomputing Research Centre are initiatives of the Australian Government, with support from the Government of Western Australia and the Science and Industry Endowment Fund. We also thank the MRO site staff. The MeerKAT telescope is operated by the South African Radio Astronomy Observatory, which is a facility of the National Research Foundation, an agency of the Department of Science and Innovation. Some of the data presented herein were obtained at Keck Observatory, which is a private 501(c)3 non-profit organization operated as a scientific partnership among the California Institute of Technology, the University of California, and the National Aeronautics and Space Administration. The Observatory was made possible by the generous financial support of the W. M. Keck Foundation. The authors wish to recognize and acknowledge the very significant cultural role and reverence that the summit of Maunakea has always had within the Native Hawaiian community. We are most fortunate to have the opportunity to conduct observations from this mountain.

This work was performed on the OzSTAR national facility at Swinburne University of Technology. The OzSTAR program receives funding in part from the Astronomy National Collaborative Research Infrastructure Strategy (NCRIS) allocation provided by the Australian Government, and from the Victorian Higher Education State Investment Fund (VHESIF) provided by the Victorian Government. We acknowledge the use of the ilifu cloud computing facility - \url{www.ilifu.ac.za}, a partnership between the University of Cape Town, the University of the Western Cape, Stellenbosch University, Sol Plaatje University, the Cape Peninsula University of Technology and the South African Radio Astronomy Observatory. The ilifu facility is supported by contributions from the Inter-University Institute for Data Intensive Astronomy (IDIA - a partnership between the University of Cape Town, the University of Pretoria and the University of the Western Cape), the Computational Biology division at UCT and the Data Intensive Research Initiative of South Africa (DIRISA). This work was carried out using the data processing pipelines developed at the Inter-University Institute for Data Intensive Astronomy (IDIA) and available at \url{https://idia-pipelines.github.io}. IDIA is a partnership of the University of Cape Town, the University of Pretoria and the University of the Western Cape. This work made use of the CARTA (Cube Analysis and Rendering Tool for Astronomy) software \cite[DOI \url{10.5281/zenodo.3377984} – \url{https://cartavis.github.io};][]{Comrie2021}. This research has made use of the NASA/IPAC Extragalactic Database (NED) which is operated by the Jet Propulsion Laboratory, California Institute of Technology, under contract with the National Aeronautics and Space Administration. This research made use of hips2fits,\footnote{\url{https://alasky.cds.unistra.fr/hips-image-services/hips2fits}} a service provided by CDS.

\bibliography{bibliography}

\end{document}